# Role of interfacial elasticity for the rheological properties of saponin-stabilized emulsions

Sonya Tsibranska,[1] Slavka Tcholakova,[1,*]
Konstantin Golemanov,[1] Nikolai Denkov,[1],
Eddie Pelan,[2] Simeon D. Stoyanov[2,3,4],

[1]*Department of Chemical and Pharmaceutical Engineering*
*Faculty of Chemistry and Pharmacy, Sofia University*
*1 J. Bourchier Ave., 1164 Sofia, Bulgaria*

[2] *Unilever R&D, Vlaardingen, The Netherlands*

[3]*Laboratory of Physical Chemistry and Colloid Science, Wageningen University, 6703 HB Wageningen, The Netherlands*

[4]*Department of Mechanical Engineering, University College London, Torrington Place, London WC1E 7JE, UK*

*Corresponding author:
Prof. Slavka Tcholakova
Department of Chemical and Pharmaceutical Engineering
Faculty of Chemistry and Pharmacy, Sofia University
1 James Bourchier Ave., 1164 Sofia
Bulgaria

Phone: (+359-2) 962 5310
Fax:    (+359-2) 962 5643
E-mail: SC@LCPE.UNI-SOFIA.BG

# Abstract

Hypothesis

Saponins are natural surfactants which can provide highly viscoelastic interfaces. This property can be used to quantify precisely the effect of interfacial dilatational elasticity on the various rheological properties of bulk emulsions.

Experiments

We measured the interfacial dilatational elasticity of adsorption layers from four saponins (Quillaja, Escin, Berry, Tea) adsorbed on hexadecane-water and sunflower oil-water interfaces. In parallel, the rheological properties under steady and oscillatory shear deformations were measured for bulk emulsions, stabilized by the same saponins (oil volume fraction between 75 and 85 %).

Findings

Quillaja saponin and Berry saponin formed solid adsorption layers (shells) on the SFO-water interface. As a consequence, the respective emulsions contained non-spherical drops. For the other systems, the interfacial elasticities varied between 2 mN/m and 500 mN/m. We found that this interfacial elasticity has very significant impact on the emulsion shear elasticity, moderate effect on the dynamic yield stress, and no effect on the viscous stress of the respective steadily sheared emulsions. The last conclusion is not trivial, because the dilatational surface elasticity is known to have strong impact on the viscous stress of steadily sheared foams. Mechanistic explanations of all observed effects are described.

# 1. Introduction.

The rheological behaviour of concentrated oil-in-water emulsions is determined by their oil volume fraction, $\Phi$, interfacial tension, $\sigma$, average radius and polydispersity of the drops [1,2]. In the emulsions with oil volume fraction above a certain critical value, $\Phi_0$, the drops deform permanently from spherical to polyhedral shape due to steric constraints. The value of $\Phi_0$ depends on the emulsion polydispersity and is in the range of $\approx 0.64$ (disordered monodisperse spheres) to $\approx 0.74$ (polydisperse samples) [1,2]. Provided that there is no adhesion between the drops, the emulsion has a purely viscous response at $\Phi < \Phi_0$ [3, 4]. On increasing the volume fraction above $\Phi_0$, the response becomes visco-elastic for both concentrated emulsions and foams [1-10].

Beside $\Phi$, $\sigma$, and mean radius, two other factors can also affect the elasticity of the concentrated emulsions and foams: (1) Attraction between the drops or bubbles [11-16] and (2) Interfacial elasticity [17-19]. It is well known that the strong drop-drop adhesion can lead to an increase of the viscosity of diluted emulsions [11,12], an increase of the elasticity of emulsions [13,14] and an increase of the viscous stress in sheared foams [15]. The effect of the surface rheological properties of the emulsion drops is still not completely clear. Dimitrova et al. [18,19] studied the rheological properties of emulsions stabilized with different proteins. They showed that higher values of the bulk elasticity of emulsions correlate with high surface elasticity. However, in their system there was also adhesion between the drops and, as a result, the authors could not clearly de-couple these two effects. Particularly high emulsion elasticity was reported also for some Pickering emulsions [20,21], which was attributed to the extremely high elasticity of the particle adsorption layers.

To clarify this issue and to de-couple the effect of surface elasticity from the effect of drop-drop adhesion, one has to employ suitable model emulsions. These emulsions should meet the following criteria: (1) To have high stability to coalescence in shear deformation. To determine reliably the bulk modulus of the emulsions, one has to be sure that there is no drop-drop coalescence during the rheological test. For example, some protein-stabilized emulsions have high surface elasticity, but they are unstable when shear deformation is applied [18,19,22]. (2) To provide a very wide range of values of the surface modulus. Such a variation ensures that a clear correlation between the surface and bulk moduli could be established. (3) To have no adhesion between the drops. As the drop-drop adhesion can also affect the bulk elasticity, there should be no strong attraction in the systems studied.

Possible candidates for appropriate model systems are the emulsions stabilized by saponins. Saponins are natural surfactants, which consist of a hydrophobic aglycone (triterpenoid or steroid) and oligosaccharide chains (1 to 3), connected to the aglycone via ester or ether bond [23-25]. Saponins are widely spread in nature – they can be found in more than 500 plant species. They are well-known for their biological activity: anti-inflammatory, anti-microbial, anti-bacterial, anti-virus, anti-cancer, and anti-tumor [24,25].

In a previous article [26] we studied the rheological properties of saponins on the oil-water interface, subject to shear deformation. The studied saponins were of triterpenoid type: Escin (ES), Tea Saponin (TS), *Quillaja* saponin (designated as QD), saponins from the species *Sapindus Mukurossi* (BSC). On the hexadecane-water interface, all saponins exhibited elasticity with different magnitudes, depending on the specific saponin. On the tricaprylin-water interface, only QD had measurable elastic modulus, while the response of the other saponins was purely viscous.

In the current article we study the surface and bulk rheological properties of emulsions stabilized with the same four saponins. Our aims are: (1) to quantify the interfacial rheological properties under dilatational deformation for hexadecane-water and sunflower oil-water interfaces, and (2) to establish a clear correlation between the bulk and interfacial elasticities of the emulsions. We showed in a previous study [27] that the dilatational and the shear surface elasticities of saponin adsorption layers, formed at the air-water interface, are related in magnitude. In addition, we found [28] that the dilatational elasticity is more relevant for the foam rheological response because the surface area of the bubbles varies under foam shear deformation, which means that the dilatational elasticity of the adsorption layers is important when studying the shear elasticity of foams or emulsions. Therefore, in the current study we first determine the surface dilatational elasticity of the saponin adsorption layers and, afterwards, we use this information to analyse the impact of the dilatational surface elasticity on the rheological properties of emulsions subject to shear deformation.

Wojciechowski and co-workers [29,30] also studied the viscoelasticity of saponin adsorption layer on the oil-water interface, but their work focused only on one saponin (*Quillaja* saponin) and they did not study the relation with the rheology of concentrated emulsions.

Apart from the more fundamental questions formulated above, the current study is important also from a more practical viewpoint. Saponins combine unique surface and biological properties, which make them a very appropriate choice as emulsion stabilizers in food, pharmaceutical and cosmetic products. However, we are aware of relatively few articles [29-41] which study their properties as emulsifiers.

The current article is organized as follows: Section 2 describes the materials and methods used. The experimental results and their discussion are presented in section 3. Sections 3.1 and 3.2 are dedicated to the interfacial tension isotherms and interfacial rheological properties, respectively. In section 3.3 we present results about the rheological properties of the bulk emulsions, and discuss the relation between the interfacial elasticity and bulk rheological properties (emulsion elasticity, dynamic yield stress and viscous stress under stead shear). The main results and conclusions are summarized in section 4.

## 2. Materials and methods.

*2.1. Materials.* We studied 5 saponin extracts of different purity. Escin, denoted as ES hereafter, is a product of Sigma (Cat. No. 50531) and contains > 95 % of the pure chemical substance aescin type II ($C_{54}H_{84}O_{23}$; CAS Number 6805-41-0) which is a monodesmosidic saponin [42]. Tea saponin extract (TS; product of Zhejiang Yuhong Import & Export Co.) contains a mixture of > 96 % monodesmodic saponins extracted from the plant *Camelia Oleifera Abel* [43]. Berry saponin concentrate (BSC) is a product of Ecological Surfactants LLC and contains ≈ 53 % saponins extracted from *Sapindus Mukurossi* [44]. Supersap (SupS) and Quillaja dry (QD) contain bidesmosidic saponins extracted from the bark of *Quillaja Saponaria Molina* tree (both extracts are products of Desert King, Chile). The saponin concentration in SupS is ≈ 91 % vs. ≈ 38 % in QD. The concentration of tanins, polyphenols and other admixtures, naturally found in the bark of *Quillaja Saponaria Molina* tree, is much higher in the QD extract, as compared to SupS [45-46]. All saponin extracts were used as received, without further purification. The saponin solutions were prepared with 10 mM NaCl as neutral electrolyte and 0.1g/l $NaN_3$ as preservative.

Table S1 in Supporting information provides information about the studied oils. Hexadecane is a product of Sigma-Aldrich and the sunflower oil was provided by Unilever, R&D (Wageningen, The Netherlands). Hexadecane (denoted as HEX below) is a saturated hydrocarbon with chemical formula $C_{16}H_{34}$. The melting point of hexadecane is 18 °C, and that is why all experiments with hexadecane were performed at 25°C. Sunflower oil (denoted as SFO) consists of triglycerides of fatty acids (linoleic 48-74%, oleic 10-14%, palmitic 4-9%, stearic 1-7%), as well as numerous admixtures - lecithin, tocopherols, carotenoids and waxes. In the experiments for characterization of the surface properties and in part of the emulsion measurements we used purified oil. The oils were purified from surface-active impurities by passing them through a glass column of chromatographic adsorbent Florisil (Supelco, Cat. No. 20280-U, PR60/100 Mesh) and silica gel 60 (Merck, Cat. No. 107734.2500) by the procedure described by Gaonkar & Borwankar [47]. To test the purity of the oils we measured the interfacial tension of the respective water-oil interface. The interfacial tension of the purified hexadecane-water interface was 52.5 ± 0.5 mN/m at 20 °C, and for purified SFO-water interface it was 30.5 ± 0.5 mN/m at 20 °C. There was no significant decrease of the surface tension over time, which indicated that there were no surface active substances dissolved in the purified oils.

*2.2. Measurement of interfacial tension (IFT) of the solutions.* To determine the IFT isotherm of TS and ESC on HEX-water and SFO-water interfaces we used the drop shape analysis (DSA) on instrument DSA100R and software DSA1 (Krüss GmbH, Germany). For measuring the IFT we formed an oil drop on the tip of a U-shaped capillary that was immersed in the aqueous saponin solution. The measurements were performed at $T$ = 25 °C.

For each concentration, at least three independent measurements were performed and the mean value with the respective standard deviation is shown in the figures and tables below.

The experimental data for the interfacial tension as a function of time, for TS and ESC on HEX-water and SFO-water interfaces, were fitted with the bi-exponential function used in Ref. [48] to describe the experimental data for air-water interface. The equilibrium interfacial tension, $\sigma_e$, was determined from the best fit to the kinetic data:

$$\sigma(t) = \sigma_e + \Delta\sigma_{1C} \exp\left(-\frac{t}{t_{1C}}\right) + \Delta\sigma_{2C} \exp\left(-\frac{t}{t_{2C}}\right) \quad (1)$$

where $t_{1C}$ and $t_{2C}$ are the characteristic relaxation times, and $\Delta\sigma_k$ are the related amplitudes of the relaxing stresses.

The experimental data for the interfacial tension isotherms were fitted by Gibbs adsorption isotherm [49]:

$$\frac{d\pi}{d\ln C} = k_B T \Gamma \quad (2)$$

where $k_B$ is Boltzmann constant, $T$ is absolute temperature, $C$ is saponin concentration in the bulk solution, $\Gamma$ is the total saponin adsorption (surface concentration), and $\pi$ is the surface pressure. From the slope of the dependence $\pi(\ln C)$ we determined the saponin adsorption at $C \approx$ CMC. The surface pressure, $\pi$, is defined as:

$$\pi = \sigma_0 - \sigma_e \quad (3)$$

where $\sigma_0$ is the interfacial tension of the pure oil-water or air-water interface.

To determine the interfacial tension of QD solutions with 0.5 wt % and 1.0 wt % at HEX-water interface, we also used the DSA method. Indeed, the error of Laplace fit to drop profile was less than 2 μm even 900 s after drop formation which evidenced that the drop shape in these systems was described adequately by Laplace equation of capillarity [50].

*2.3. Measurement of the surface dilatational modulus by capillary pressure tensiometry.* Capillary pressure tensiometry (CPT) is a technique for measurement of the surface rheological properties of liquid-fluid interfaces [51-53]. In this method we form a semi-spherical drop on the tip of a capillary. The interfacial area of the drop is subjected to sinusoidal deformation (expansion, contraction). The variation of the pressure, $\Delta P$, in the drop is measured via a pressure transducer and the related variation in the surface tension, $\Delta\sigma$, is determined. The amplitude in the variations of the interfacial tension, divided by the amplitude of the variations of surface area, gives direct information for the surface modulus, see Refs. [48,51-53]. The measurements were performed on DSA30 instrument (Krüss, Germany). The adsorption layer was left to equilibrate for 15 minutes prior to the measurement. The frequency of oscillation was 0.1 Hz. The temperature was 25 °C. For each system, at least three independent measurements were performed and the mean value with its standard deviation is presented in the figures and tables below.

*2.4. Preparation of the emulsions.* All emulsions were produced at room temperature (≈ 25 °C). We prepared the premix in a planetary mixer (Kenwood Chef Premier KMC 560; 1000 W) by the following procedure: first we added the solution of saponin in the container, after that we added slowly the oil (sunflower oil or hexadecane) for period of 4 min at 1.5 rps speed. After the total amount of oil was added, we homogenized the mixture for additional 4 min at the same speed (1.5 rps). After that we transferred the premix in the colloid mill (Magic Lab; MK; U078310). In our experiments the gap was 398 μm and the rate of rotation was 10 000 rpm. The homogenizer was connected to a pump (ISMATEC; MCP-CPF Process IP65; Pump head - FMI212/QP.Q2.CSC/9004) which facilitates the emulsion flow. The pump operated at 500 rpm rate of rotation. The total concentration of saponin in the experiments was 1 wt. %. The total volume of the liquid (water + oil) was fixed at 200 ml.

To check for the effect of surface active admixtures present in the oils, we repeated the main emulsion experiments with purified oils (the same as those used to characterise the interfacial properties of the saponin solutions). The obtained experimental results for the emulsions prepared by purified and non-purified oils did not show any statistically significant difference. Therefore, we present below the average results from both series of experiments.

For each system, at least three different emulsions were prepared and characterized independently. Below we show the mean values with their standard deviations.

*2.5. Microscopic observation of the emulsions and the solutions.* We performed optical observations in transmitted light of the emulsion drops using optical microscope Axioplan (Zeiss, Germany), equipped with objective Zeiss Epiplan 50x. The emulsion samples were diluted with saponin solution. Afterwards the diluted samples were placed in a capillary tube with rectangle cross-section, made of borosilicate glass.

The drop size distribution was measured with microscope Axio Imager M2m (Zeiss, Germany). The same rectangle capillary with emulsion sample and transmitted light were used. The microscope was equipped with a sample holder, which allowed automatic scan of the samples, during which the microscope took a series of pictures of the emulsion drops. The microscope was equipped with software Axio Vision which processed the recorded images, determining the radius of each drop and calculating the mean surface-to-volume radius, $R_{32}$:

$$R_{32} = \sum_i R_i^3 \Big/ \sum_i R_i^2 \qquad (4)$$

where the sum is taken over all measured drops in a given sample. At least 1000 drops were measured for each emulsion. For most of the samples, more than 20000 drops were measured.

***2.6. Characterization of the rheological properties of the emulsions.*** The rheological properties of the emulsions were studied using rotational rheometer (Bohlin Gemini, Malvern Instruments, UK). We used parallel plates geometry with diameter, $d$ = 40 mm. On both plates we glued sandpaper (grade P1500) to suppress the possible wall slip.

We performed two types of rheological tests:

(A) <u>Oscillatory deformation (amplitude sweep)</u>. The strain amplitude, $\gamma_A$, was varied logarithmically from 0.05 to 50 % at fixed frequency (1 Hz). We measured the elastic, $G`$, and viscous, $G``$ moduli of the emulsion, as a function of strain amplitude, $\gamma_A$. The elasticity of the emulsions was characterized by the value of $G`$ in the plateau region of the plot $G`$ vs. $\gamma_A$ (see section 3.4 for more details).

(B) <u>Steady shear deformation.</u> In this test, the shear rate, $\dot{\gamma}$, was varied logarithmically from 0.5 s$^{-1}$ to 50 s$^{-1}$ for 180 s and the shear stress was recorded, $\tau = \tau(\dot{\gamma})$. The results were described with the Herschel-Bulkley rheological model:

$$\tau = \tau_0 + k\dot{\gamma}^n \qquad (5)$$

where $\tau_0$ is the yield stress of the emulsions, $k$ is the consistency and $n$ is the power-law index. $\tau_0$, $k$ and $n$ were determined from the best fit of the experimental results.

These experiments were performed in the following manner. First we loaded a sample of emulsion at gap between the plates of 2000 μm, and performed test (A). Then we performed test (B) at the same gap. Afterwards we decreased the gap down to 1500 μm and again performed test (A). In the majority of the experiments there was no difference in the results for $G`$ from the first test (before steady shear) and the second test (after steady shear). In some of the systems $G`$ measured in the second test was significantly lower, as a result of the shearing of the sample. These results are discussed in more detail in the relevant section 3.4.

**3. Experimental results and discussion.**

***3.1. Interfacial tension isotherms of saponins.*** Figure 1 presents the interfacial tension isotherms of Escin and Tea Saponin on the air-water and oil-water (Hexadecane; SFO) interfaces. To compare the obtained results for interfaces with different $\sigma_0$ (surface tension of the surfactant-free interface) we plot the surface pressure as a function of the concentration of saponin. These results are fitted with the Gibbs adsorption isotherm, equation 2. From the best fit we determined the saponin adsorption at CMC, $\Gamma_{CMC}$. Table 1 presents results for the adsorption of saponin and the area per molecule at CMC on air-water or oil-water interfaces.

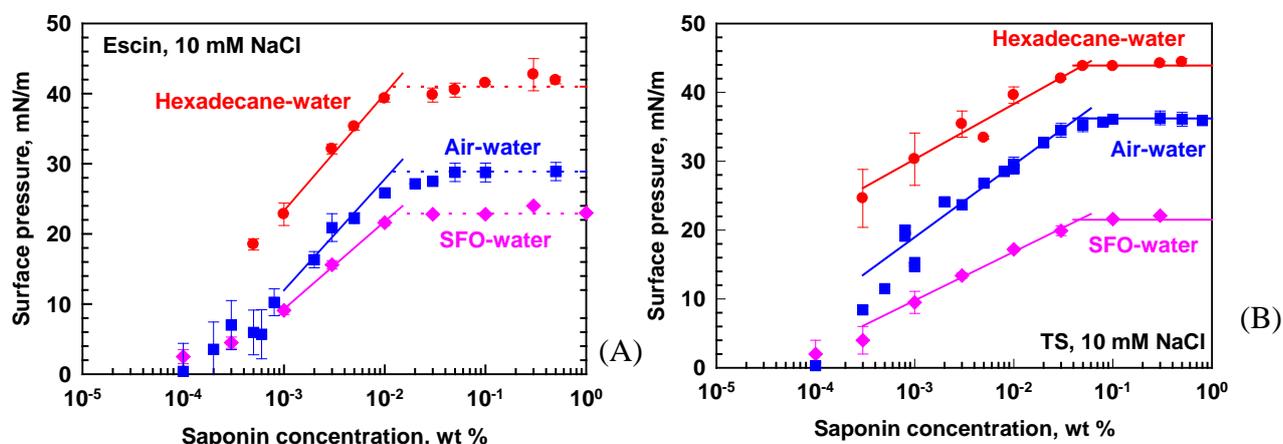

**Figure 1.** Surface pressure vs. saponin concentration for oil-water and air-water interfaces: (A) Escin; (B) Tea Saponin. CMC for Escin is $0.013 \pm 0.002$ wt % and for TS it is $0.05 \pm 0.01$ wt %. The adsorption at CMC is determined from the linear dependence of the surface pressure on the logarithm of saponin concentration, $\lg C$. For Escin, the range of $C$ between 0.001 and 0.01 wt % is used, and for TS, the range of $C$ between 0.001 and 0.03 wt % is used to determine adsorption. The data for air-water interface are taken from Pagureva et al. [27].

**Table 1.** Adsorption and area-per-molecule in the surfactant adsorption layers at CMC for oil-water and air-water interfaces.

| Interface | $\Gamma_{CMC}$, μmol/m$^2$ | | $A_{CMC}$, nm$^2$ | |
|---|---|---|---|---|
| | Escin | Tea Saponin | Escin | Tea Saponin |
| Air-Water* | $2.9 \pm 0.2$ | $1.9 \pm 0.2$ | $0.57 \pm 0.05$ | $0.87 \pm 0.09$ |
| HEX-Water | $2.9 \pm 0.2$ | $1.4 \pm 0.2$ | $0.57 \pm 0.05$ | $1.16 \pm 0.12$ |
| SFO-Water | $2.2 \pm 0.2$ | $1.2 \pm 0.2$ | $0.76 \pm 0.07$ | $1.38 \pm 0.23$ |

*- data taken from Pagureva et al. [27]

The data show that both saponins form less densely packed adsorption layers on the sunflower oil-water interface, compared to the air-water interface. The lower density could be explained by molecules of triglycerides intercalated between the saponin molecules [26]. Similar effect is observed with Tea Saponin on water-hexadecane interface. The adsorption is also lower compared to the air-water interface. However Escin exhibits similar adsorption on the two interfaces (hexadecane-water and air-water). In other words, the hexadecane molecules do not disrupt the structure of the dense escin adsorption layer detectably.

We could not measure the interfacial tension isotherms of QD or BSC on the sunflower oil-water and water-hexadecane interfaces by using drop shape analysis, as the adsorption layers had extremely high elasticity and rigidity. As mentioned above, the drop shape analysis is not a suitable method for the measurement of the surface tension of such solutions. Up to $5\times10^{-4}$ wt. % saponin concentration, we were able to describe adequately the profile of the drops with the Laplace equation. However, at higher saponin concentrations the quantitative parameter which determines the quality of the fit (FE) [54] increases significantly which indicates that the surface tension cannot be determined reliably, because additional factors affect the shape of the oil drop, apart from the gravity and surface tension. Similar effect was observed by Alexandrov et al. 2012, for adsorption layers of hydrophobin (type II) on air-water interface [50]. Hydrophobin is a protein, isolated from fungi [55], which is known to form rigid adsorption layers observed that at given surface tension the fit error started increasing steadily with the subsequent decrease of the surface tension. This effect was attributed to the formation of an elastic protein skin on the drop surface. The same process is observed in our experiments. At saponin concentration $> 5\times10^{-4}$ wt. % the elasticity of the layers increases significantly, which is indicated by the increasing value of the fit error. Illustrative results for the interfacial tension and fit error vs. time for BSC (water-hexadecane) are presented on Figure S1 in Supplementary material. One sees that the fit error for most of the studied concentrations is above 0.5 µm, which is the threshold value above which the solidification of adsorption layer of hydrophobin is observed [50]. For irreversible adsorption, such as the adsorption of hydrophobins, saponins or solid particles on fluid interfaces, the surface tension isotherms cannot be determined by the conventional methods, because no complete thermodynamic equilibrium is established between the surfactant solution and the interface in the timeframe of the experiments.

Therefore from this series of experiments we can conclude that the solidification of the adsorption layers is observed for QD and BSC layers on SFO-water and on hexadecane-water interfaces, whereas for TS and escin the adsorption layers with larger area per molecule are formed on SFO-water interface, as compared to hexadecane-water and air-water interfaces.

### *3.2. Interfacial rheological properties of saponins.*

Figures 2 present illustrative images for drops with very high surface elasticity. Figure 2A shows a drop of BSC solution in sunflower oil at rest. When the solution is sucked out of the drop, wrinkles are formed on the drop surface, see Figure 2B. Similar wrinkles are already reported in the literature for QD on air-water interface [48] as well as for polymeric Janus Particles [56], polymer/carbon nanotube [57] and Asphaltene-laden interfaces [58]. In Figure 2C we present a picture of sunflower oil drop in solution of Quillaja saponin. When we increase the volume of the drop, the radius of the drop does not increase, but instead an oil neck is ejected from the capillary Figure 2D. The drop moves like a solid body, and does not relax spontaneously, as we would expect from a conventional capillary system with fluid interface. It is important to note that this effect is observed only if the solution seeps between the oil and the wall of the capillary. In other words in order for the solid shell to be formed, the oil and the saponin solution must have been in contact, prior to the oil ejection. These experimental results unambiguously showed that QD and BSC formed solid adsorption layers on SFO-water interface.

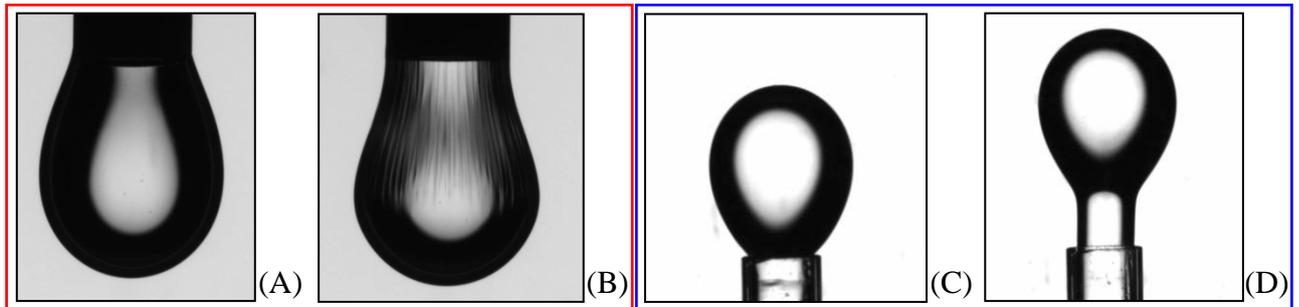

**Figure 2.** Drops of (A, B) 0.5wt% aqueous solution of BSC in sunflower oil and (C, D) sunflower oil in 0.5wt% solution of QD. (A and C) Equilibrium shape; (B) Water is sucked out of the drop and wrinkles form on the interface; (D) Oil is pushed into the drop, the whole surface (viz. the drop and the neck) moves like a rigid body.

As explained above, CPT is the suitable method to study systems with pronounced surface elasticity. However, even in the CPT set-up we observed non-Laplacian shapes with some of the saponin systems studied, see Figure 3. Figure 3A presents a drop of sunflower oil in solution of BSC in equilibrium. Part of the oil is sucked inside the capillary and the surface of the drops wrinkles, while a spherical segment shape is observed (trapeze-like in a lateral profile). Similar shape was observed previously with oil-water interfaces, covered with polystyrene particles [59]. Both QD and BSC form a solid shell on the SFO-water interface with extremely high surface elasticity (rigidity) which cannot be quantified via CPT.

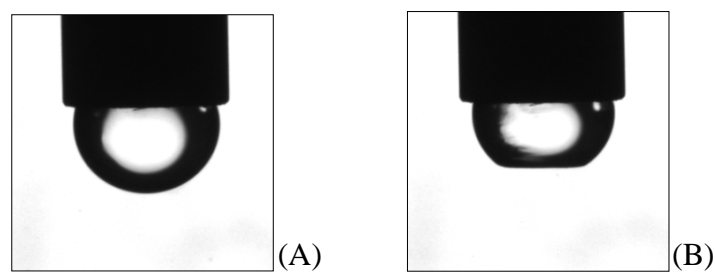

**Figure 3.** Drop of sunflower oil in 1 wt% solution of BSC. Semi-spherical drop is formed on the tip of a capillary. (A) Initial equilibrium configuration; (B) Oil is sucked out of the drop, wrinkles form on the interface and the drop acquires the shape of a spherical segment.

For the other systems we were able to measure the interfacial dilatational elasticities by CPT and the experimental results are presented in Table 2. First we compare the experimental results obtained at air-water and oil-water interfaces. To do this we measured the interfacial dilatational elasticity for saponin solutions with concentration of 0.5 wt %. The comparison shows that for Escin, Tea saponin and Supersap the dilatational surface modulus for oil-water interfaces are lower as compared to those measured with air-water interfaces. Similar effect was observed with saponin adsorption layers subject to shear deformation [26]. The lower shear and dilatational surface moduli are related to the intercalation of oily molecules in between the saponin molecules for Escin on SFO-water and for TS on Hexadecane-water and SFO-water interfaces, as evidenced by the larger area-per-molecule measured at oil-water interfaces, see Table 1.

On the other hand, the area per molecule for Escin molecules is similar for air-water and Hexadecane-water interfaces, while the dilatational and shear elasticities are much lower for Hexadecane-water interface, which is probably related to the effect of hexadecane molecules on the orientation of the escin molecules in the adsorption layer. In our previous studies [60,61] we performed classical atomistic dynamics simulations of adsorbed escin molecules on the water surface. The results revealed that long-range attraction between the hydrophobic aglycones, combined with intermediate dipole-dipole attraction and short-range hydrogen bonds between the sugar residues in escin molecules control the viscoelastic properties of escin adsorption layers on air-water interface. When oil molecules are able to accommodate between the neighbouring molecules in the adsorption layer (e.g. upon shear or upon dilatational of the surface), the presence of hexadecane molecules in the adsorption layer will decrease the attraction between the aglycones and will decrease the elasticity of the layer as a whole, even at the same surfactant adsorption.

In the case of TS, both hexadecane and SFO decrease drastically the surface modulus from ≈ 500 mN/m down to < 11 mN/m which is consistent with the fact that the adsorption of TS is lower on oil-water interfaces when compared to air-water interface. The same effect of the oil phase is observed also with Supersap. However, in contrast to TS, the modulus on the SFO-water interface is measurably higher compared to hexadecane-water.

The effect of the oil on the adsorption layers of QD and BSC is qualitatively different in comparison to the other three saponin extracts. The elastic modulus of QD and BSC on the SFO-water interface is higher, compared to the air-water interface which indicates better packing of the adsorbed molecules on SFO-water interface – this is unexpected effect which deserves more detailed consideration presented below.

Let us compare the surface properties of Supersap and QD. As mentioned above, both extracts contain saponins from *Quillaja Saponaria Molina* plant, but Supersap has much higher saponin content: ≈ 91 vs. ≈ 38.3 %. The interfacial modulus of Supersap is significantly lower at oil-water interface, whereas the modulus of QD is either not affected (hexadecane, 0.5 wt. % saponin) or it is even higher (SFO). Therefore, we conclude that the admixtures present in the QD extract affect significantly the surface rheological response of the adsorption layer. In the work by Wojciechowski [29] the effect of the hydrophobic phase on the properties of adsorption layers of the saponin extract from *Quillaja Saponaria Molina* is also studied. These authors showed that the surface dilatational elasticity of this triterpenoid saponin decreases in the order: air–water > tetradecane–water > olive oil–water interface. No explanation for the observed dependence was proposed. However, the authors showed also that the extracts of *Quillaja Saponaria Molina* may differ significantly in composition and in their surface properties, depending on their producer.

As can be seen from Table 2, there is a significant irreproducibility in the measurement of the surface modulus of BSC on the hexadecane-water interface. This effect can be explained as follows. Studies in the literature indicate that the BSC extract contains mono- and bi-desmosidic saponins, the monodesmosides being solubilized in the micelles of the bidesmoside fraction [64]. It is possible that the more hydrophobic components are transferred into the oily phase during the measurement, the interfacial layer is out of equilibrium and, as a result, we cannot reach a steady reading.

It is interesting to note that the increase of the concentration of saponin from 0.5 to 1.0 wt. % leads to a significant increase of the elastic modulus of ES, Supersap and QD layers on the hexadecane-water interface. At concentration of saponin of 1.0 wt. % the modulus of ES and Supersap is still lower compared to the air-water interface, but for QD it is almost 2 times higher. Such an effect is not expected as 0.5 wt. % is much higher than the CMC value of all of the three surfactants. Therefore, we would not expect any significant change in the equilibrium structure of the layer coming from the increase of the saponin concentration in this concentration range. This effect could be attributed to the increased concentration of saponin molecules in the sub-surface layer. It is possible that due to the higher population of saponin molecules near the interface, the disruption of the saponin adsorption layer by the oil molecules (upon layer deformation) is suppressed. In the case of sunflower oil we do not observe any measurable effect of the concentration of saponin.

**Table 2.** Interfacial dilatation moduli of saponin adsorption layers on the air-water and oil-water interfaces. Period of oscillation, $T = 10$ s; Amplitude of deformation, $1 \pm 0.2$ %. Measurements performed with 0.5 or 1.0 wt. % saponin solutions.

| Saponin | Interfacial Dilatational elasticity, mN/m | | | | |
|---|---|---|---|---|---|
| | Air-water* | HEX-W | | SFO-water | |
| | 0.5 wt. % | 0.5 wt. % | 1.0 wt. % | 0.5 wt. % | 1.0 wt. % |
| ES | 1100 ± 110 | 85 ± 15 | 370 ± 30 | 40 ± 10 | 40 ± 5 |
| TS | 480 ± 110 | 7 ± 6** | 11 ± 0.2** | 2 ± 0.1** | 2 ± 0.4** |
| Supersap | 300 ± 30 | 25 ± 3 | 50 ± 4 | 140 ± 50 | 156 ± 35 |
| QD | 260 ± 1 | 275 ± 30 | 510 ± 40 | "solid" shell | |
| BSC | 230 ± 7 | 150 ± 50*** | 140 ± 60*** | "solid" shell | |

*-Results taken from ref. 27.
**-Measurement performed with Oscillating Drop Method.
***-Scattered results, see text for explanations.

We can summarize the most important results for the rheological properties of the saponin layers in the following way. With the exception of Tea Saponin, all studied saponins exhibit high elasticity on the oil-water interface. The surface modulus of ES, Supersap and TS is significantly lower on the oil-water, compared to the air-water interface. The lowest moduli are measured with sunflower oil for TS and ES. The effect of the oil is attributed to changes in the structure of the adsorption layer, induced by intercalated oil molecules. In the case of TS (hexadecane, SFO) and ES (SFO) the lower elasticity correlates with higher surface area per molecule. The elastic modulus of QD and BSC on the sunflower oil-water interface is extremely high and the adsorption layers is rigid - the value of the surface elasticity cannot be measured due to the formation of wrinkles or solid shell on the interface. This effect is most probably due to the admixtures present in these extracts. Increasing the saponin concentration form 0.5 to 1.0 wt. % was accompanied by an increase of the elasticity on the hexadecane-water interface for layers of the ES, Supersap and QD.

Another important result is that we can vary the interfacial elasticity between 2 and 500 mN/m by using different saponin-oil pair which allows us to study quantitatively the relation between the drop surface elasticity and emulsions rheological properties.

*3.3. Effect of the interfacial properties on the shape of the emulsion drops.*

Figure 4 presents illustrative images of emulsions drops of sunflower oil, stabilized by QD or BSC and drops of hexadecane, stabilized by TS. The concentration of saponin was 1 wt. %. The sample is obtained after dilution in the respective saponin solution of an initial, more concentrated emulsion with oil volume fraction of 75 %. One sees that both QD and BSC stabilize drops with non-spherical shape, which does not relax for over 30 days. This is related to the extremely high surface modulus of the saponin adsorption layers. Such irregular drop shapes were observed in the literature also for other surface-active species with high surface elasticity: solid particles [59, 63-65], polysaccharides [66], proteins [67]. We remind that the other saponins, with the exception of TS, also exhibit high elasticity with SFO and hexadecane. However, in this case we do not observe any irregularity in the drop shape at the same concentration of saponin (1 wt. %). Probably, there was a deviation from the spherical shape for these systems initially, but the shape relaxed towards the spherical one over time. We did not observe any drop-drop aggregation in the hexadecane and SFO emulsions, stabilized by QD, BSC or ES. Therefore, for these emulsions we can clearly rule out the adhesion between the drops as a possible factor which could affect the bulk emulsion elasticity. In the case of Tea Saponin we observed some drop aggregates in the emulsions of hexadecane (1 wt. % saponin), Figure 4C.

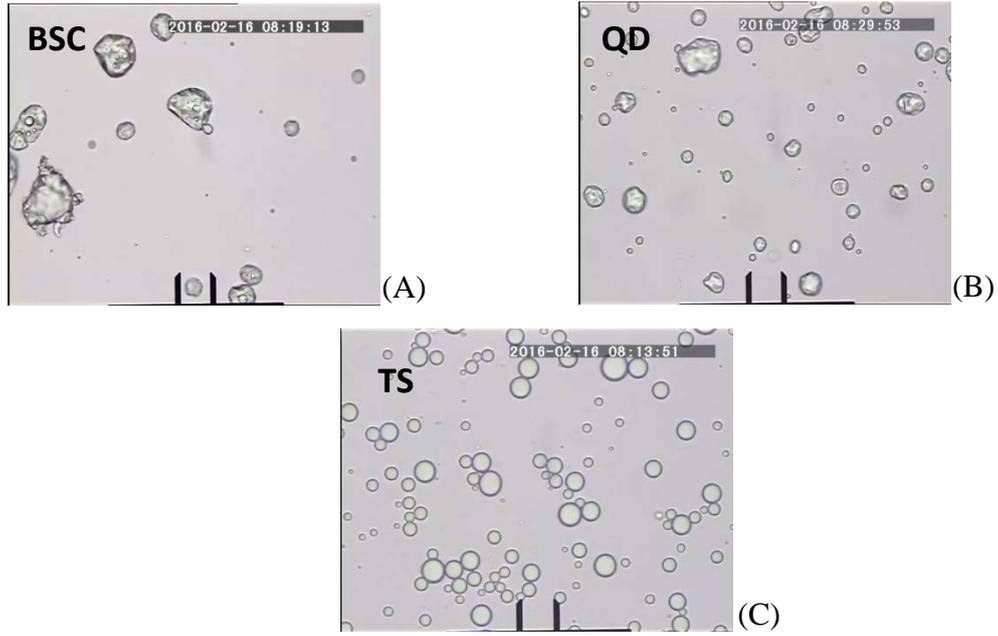

**Figure 4.** Emulsion drops of sunflower oil in solutions of (A) BSC or (B) QD. (C) Hexadecane drops stabilized by TS. Concentration of saponin in all solutions: 1 wt. %. Oil volume fraction in the original emulsion: 75 %. The distance between the vertical black bars in the bottom of these images is 20 μm.

### *3.4. Effect of the interfacial properties on the rheological properties of the emulsions.*

We start this section by discussing the results for the elastic modulus of the bulk emulsions, obtained in oscillatory deformation (amplitude sweep). Figure 5 presents illustrative results for the elastic modulus, $G`$, and viscous modulus, $G``$, of hexadecane emulsion stabilized by QD, as a function of the amplitude of deformation, $\gamma_A$. It can be seen that $G`$ does not change significantly with $\gamma_A$ at small deformations (plateau region). At higher values of $\gamma_A$, $G``$ goes through a maximum, while $G`$ starts decreasing. In the discussion below we consider the value of the elastic modulus in the plateau region ($\gamma_A < 0.2$ %), which is designated as $G`$. The elasticity of the regular emulsions is relatively well understood. As "regular" we mean emulsions in which no adhesion between the drops is present and the interfacial elasticity is negligible. Princen and Kiss [68] and Mason et al. [69] proposed the following expressions for the elastic modulus of such emulsions:

$$\tilde{G}`_P = \frac{G`R_{32}}{\sigma} = 1.77\Phi^{1/3}(\Phi - \Phi_0); \quad \Phi_0 = 0.72 \quad \text{(Princen, polydisperse systems) (6)}$$

$$\tilde{G}`_M = \frac{G`R_{32}}{\sigma} = 1.7\Phi(\Phi - \Phi_0); \qquad \Phi_0 = 0.64 \qquad \text{(Mason, monodisperse systems) (7)}$$

In both cases the authors measured the elasticity of a series of emulsions with different $\Phi$, $\sigma$, $R_{32}$ and described their results with the semi-empirical equations (6) and (7).

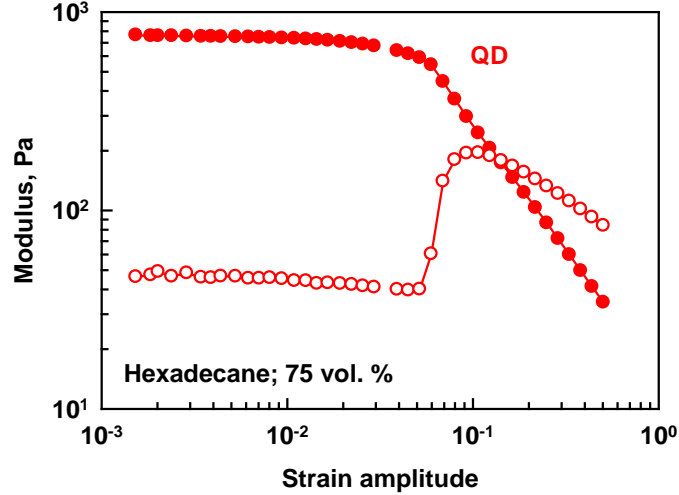

**Figure 5.** Elastic (full symbols) and viscous (empty symbols) moduli of hexadecane emulsions, stabilized by QD. Oil volume fraction: 75 vol. %. Frequency of deformation: 1 Hz.

Figure 6 presents results for the elastic modulus of the emulsions of SFO-in-water or hexadecane-in-water, stabilized by different saponins and having different oil volume fractions (75, 80, 85 vol. %). In several cases we were unable to produce stable oil-in-water emulsions (e.g. Escin + hexadecane; $\Phi_{OIL}$ = 75, 80, 85 %) and, therefore, we do not present data for the elasticity of these emulsions. The results we discuss here are obtained before the pre-shear, from the first amplitude sweep test. As expected, the emulsion modulus increases with increasing the oil volume fraction. We observed also that the elasticity varied significantly depending on the saponin type. At least partially this effect can be attributed to variations in the surface tension and in the size of the emulsion drops in the different systems. To account for these two factors, we scaled the measured emulsion elastic modulus with the drop capillary pressure: $\tilde{G}` = G`R_{32}/\sigma$. The results for the obtained dimensionless elastic modulus are presented in Figure 6 C,D.

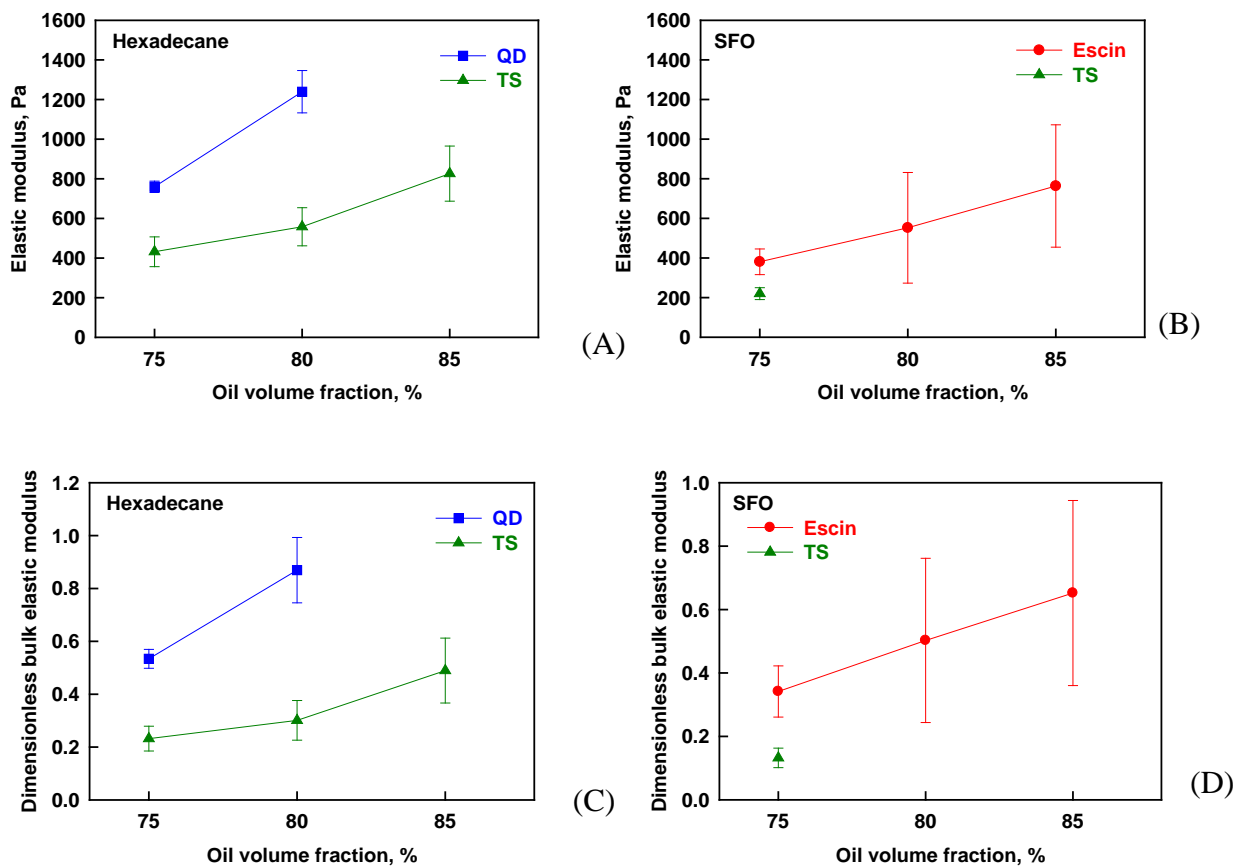

**Figure 6.** (A,B) Elastic modulus and (C,D) Dimensionless elastic modulus vs. oil volume fraction for emulsions stabilized by different saponins. Concentration of saponin: 1 wt. %. Saponins: Tea saponin (green triangles); Escin (red squares); QD (blue circles). Oil phase: (A,C) Hexadecane; (B,D) Sunflower oil.

To account for the effect of oil volume fraction, we re-plotted the experimental data shown in Figure 6 as a ratio of the dimensionless bulk elastic modulus and the predicted values by eq. (7). The obtained results, as a function of the oil volume fraction, are shown in Figure S2. As can be seen from Figure S2, eq. (7) accounts properly for the effect of oil volume fraction. That is why we determined the average value of the normalized elasticity from all prepared emulsions for given saponin-oil pair. The obtained results are plotted in Figure 7, as a function of the measured interfacial dilatational modulus for the same system. The attempt to use eq. (6) for accounting the effect of oil volume fraction led to much worse comparison – instead of having constant value for given system (as a function of oil volume fraction), we obtained normalized elasticities which decreased significantly with the increase of the oil volume fraction. In other words, eq. (6) underestimated the emulsion elasticity at low oil volume fractions for the systems studied. Therefore, only eq. (7) is used in the further analysis.

The normalized elasticity as a function of the dilatational surface elasticity is shown in Figure 7. One sees that the emulsion elasticity is well described by eq. (7) when the surface dilatational elasticity is below 5 mN/m. When the bulk elasticity increases up to 10 mN/m, however, slight deviation from the predicted value is observed and the measured elasticity is ≈ 50 % higher as compared to the predicted one. The further increase in the surface elasticity up to 500 mN/m leads to ≈ 4 times increase in the emulsion elasticity. The proposed relation between the normalized bulk elasticity and the surface elasticity can be used to determine the value of bulk elasticity from the interfacial tension, oil volume fraction and surface dilatational elasticity using the following relation (where $E_D$ is expressed in mN/m):

$$G` = \begin{cases} 1.7 \dfrac{\sigma}{R_{32}} \Phi(\Phi - 0.64) & E_D < 5 \text{ mN/m} \\ \\ 1.7 \dfrac{\sigma}{R_{32}} \Phi(\Phi - 0.64)(1.5 \lg E_D) & E_D > 5 \text{ mN/m} \end{cases} \qquad (8)$$

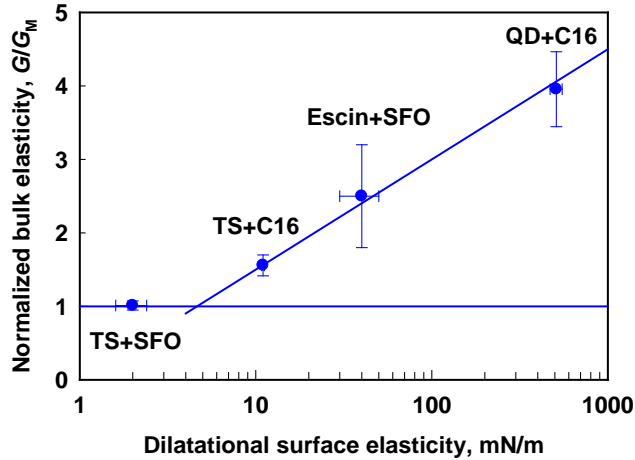

**Figure 7.** Normalized bulk elasticity as a function of the dilatational surface elasticity in linear-log scale.

The effect of surface dilatational elasticity on the bulk emulsion shear elasticity can be explained in the following manner. When the emulsion is in equilibrium, and it is not subject to mechanical agitation, the drop has a given interfacial area, $s_i$. The value of $s_i$ is defined mostly by the oil volume fraction. When shear is applied, the drops are deformed, and the interfacial area increases to a value $s > s_i$. This results in an increase of the interfacial energy of the drops and to increase of the emulsion energy. Macroscopically that is registered as bulk elasticity [28].

The elastic modulus, $G`$, is a characteristic of the elasticity of the emulsions. Another characteristic which is related to the deformation of the droplets before emulsion flowing is the emulsion yield stress, $\tau_0$, which we determined from the shear ramp experiments. The experimental results $\tau(\dot{\gamma})$ were described by the Herschel-Bulkley rheological model, see eq. (5) above. Similarly, to the elastic modulus, the yield stress can be scaled with the drop capillary pressure, $\tilde{\tau}_0 = \tau_0 R_{32} / \sigma$. Mason et al. [69] derived the following empirical expression for $\tilde{\tau}_0$:

$$\tilde{\tau}_{0,M} = \frac{\tau_0 R_{32}}{\sigma} = 0.51(\Phi - 0.62)^2 \qquad (9)$$

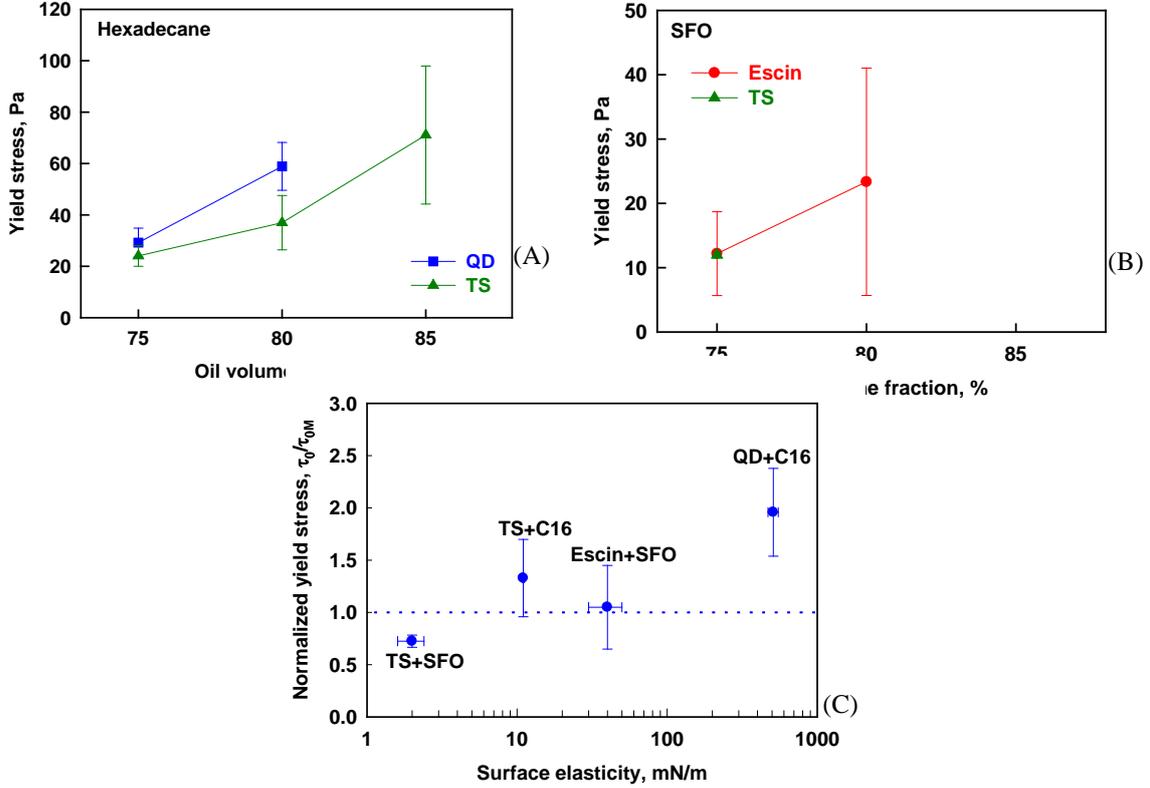

**Figure 8.** (A,B) Yield stress vs. oil volume fraction of emulsions stabilized by different saponins: Tea saponin (green triangles); Escin (red squares); QD (blue circles). Concertation of saponin: 1 wt. %.. Oil phase: (A) Hexadecane; (B) Sunflower oil. (C) Normalized yield stress, as a function of the dilatational surface elasticity.

Figure 8A,B presents experimental results for the yield stress vs. oil volume fraction for emulsions, stabilized by different saponins. Figure 8C presents the normalized value of the measured yield stress with the prediction of eq. (9). One sees that the effect of surface elasticity is much less pronounced for the emulsion yield stress, as compared to its effect on emulsion bulk elasticity. Despite the fact that both $G`$ and $\tau_0$ characterize the emulsion rheological properties before sliding the droplets with respect to each other, $G`$ is much more sensitive to the increased interfacial elasticity as compared to the yield stress. This is probably related to the fact that we determine $\tau_0$ as the minimal shear stress which has to be applied in order to induce irreversible emulsion deformation. Note, however, that in our measurements we determine the so-called dynamic yield stress [70, 71] is determined the best fit to the experimental results $\tau = \tau(\dot{\gamma})$. During this experimental test $\tau > \tau_0$ and the emulsion flows, while the surfaces of the emulsion drops are involved in oscillatory dilatational deformation.

This drop surface deformation can result in a decrease of the drop surface modulus and, respectively, of the measured yield stress. In other words, the used steady shear test probably modifies the rheological properties of the drop surfaces and of the emulsion itself. In contrast, the elastic modulus, $G`$, is measured at very small deformations ($\approx 0.05$ %). The mechanical agitation is much milder and does not affect the surface modulus of the drops. That is why $G`$ is more indicative for the effect of the surface elasticity on the bulk elastic properties of emulsions in rest.

Other two important rheological characteristics of the emulsions involved in steady shear deformation are the power-law index, $n$, and the consistency, $k$, see eq. (5) above. In our previous study we showed that the interfacial dilatational elasticity has significant impact on both $n$ and $k$ for sheared foams [72,73]. It was shown that for foams with dilatational surface elasticity > 50 mN/m the flow index decreases from $\approx 0.5$ to $\approx 0.2$ and the dimensionless viscous stress is much higher at the same capillary number [73]. To check whether a similar effect is observed for emulsions studied here, we determined the values of the flow index for different emulsions. We found that for all studied systems $n \approx 0.47$, which is very close to the theoretically predicted value [74,75] based on the assumption that the viscous dissipation occurs in the films formed between the sliding droplets. No explicit dependence on the interfacial dilatational elasticity was observed, see Figure 9 which differs significantly from the results reported for foams. This result is probably related to the fact that the studied saponin layers have predominantly elastic response, whereas the change in flow index for foams was explained with the high viscous surface modulus, measured with the foam systems.

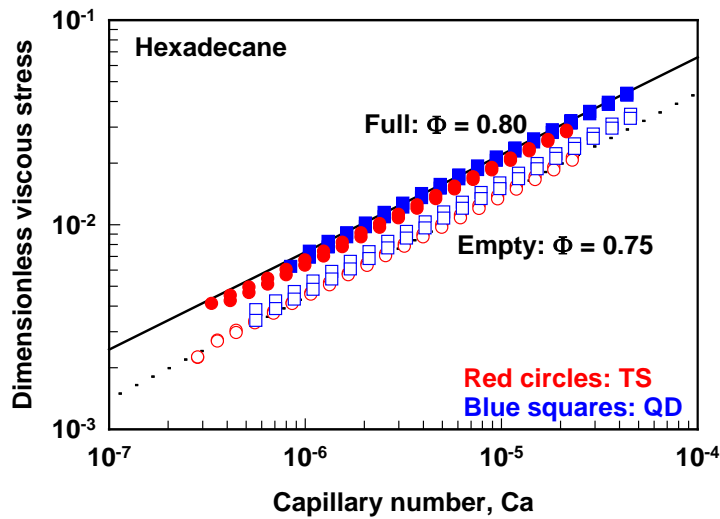

**Figure 9.** Dimensionless viscous stress, as a function of the capillary number Ca for hexadecane emulsions stabilized by different saponins with different surface elasticities. Concertation of saponin: 1 wt. %. Volume fraction: 0.80 (full symbols); 0.75 (empty symbols). Saponins: TS (red circles); QS (blue squares).

## 4. Main results and conclusions.

In the current study we characterized the interfacial properties of saponin adsorption layers (from Escin, Tea saponin, Quillaja saponin, and Berry Saponin concentrate) formed at oil-water interfaces. We performed experiments with two oils having very different molecular structures: hydrocarbon (hexadecane) and triglyceride (sunflower oil). We measured the interfacial tension isotherms of the saponins and the interfacial dilatational moduli. The obtained results are compared with the surface properties of the same saponins for air-water interface. In parallel, we studied the bulk rheological properties of the emulsions, stabilized by these saponins, with the major aim was to clarify the relation between the interfacial and bulk emulsion viscoelasticities. The main results and conclusions can be summarized as follows.

The increase of the interfacial dilatational elasticity has significant impact on the drop shape and on the emulsion rheological properties, measured under steady shear deformation. The formation of solid adsorption layers on the SFO-water interface, when BSC and QD saponins are used, leads to formation of emulsion drops with non-spherical shape which does not relax for more than 30 days. The increase of the interfacial dilatational elasticity from 10 to 500 mN/m increases the dimensionless elasticity of the bulk emulsions by more than 4 times, the dimensionless yield stress by $\approx$ 2 times, and has no effect on the viscous stress of the concentrated emulsions. The later result is very different from the results reported in the literature for foams [72,73] and can be explained with the predominantly elastic response of the studied saponin adsorption layers.

The current study complements a previous work which showed that the surface elasticity of saponin adsorption layers have significant impact on the rate of bubble Ostwald ripening in saponin-stabilized foams [76]. Currently we work to reveal the effect of saponin adsorption layers on the rheological properties of foams and on several other foam and emulsion properties. Another direction of research, promising but not explored yet, is to use saponins to test the effect of drop-drop aggregation on the rheological response of foams and emulsions. Thus we see that the saponins can be used as a very convenient model system to quantify the effect of interfacial viscoelasticity on various foam and emulsion properties. Taking into account also the bio-activity of various saponins and their related potential applications as surfactants and bio-active ingredients [77-79], we expect that this opportunity will spark numerous studies in the following years.

**Acknowledgements**
The authors are grateful to PhD Student Nevena Pagureva for performing the interfacial tension measurements.

**Funding:** This work was supported by Unilever R&D Vlaardingen, the Netherlands, and by Operational Program "Science and Education for Smart Growth" 2014-2020, co-financed by European Union through the European Structural and Investment Funds, Grant BG05M2OP001-1.002-0012 "Sustainable utilization of bio-resources and waste of medicinal and aromatic plants for innovative bioactive products". S. Tsibranska is grateful to Operational program "Science and Education for Smart Growth", project BG05M2OP001-2.009-0028 for the financial support for her scientific visit to Unilever R&D Vlaardingen, the Netherlands.

.

**Graphical abstract**

Role of interfacial elasticity for the rheological properties
of saponin-stabilized emulsions
by
Sonya Tsibranska,[1] Slavka Tcholakova,[1*] Konstantin Golemanov,[1]
Nikolai Denkov,[1] Eddie Pelan,[2] Simeon D. Stoyanov[2,3,4]

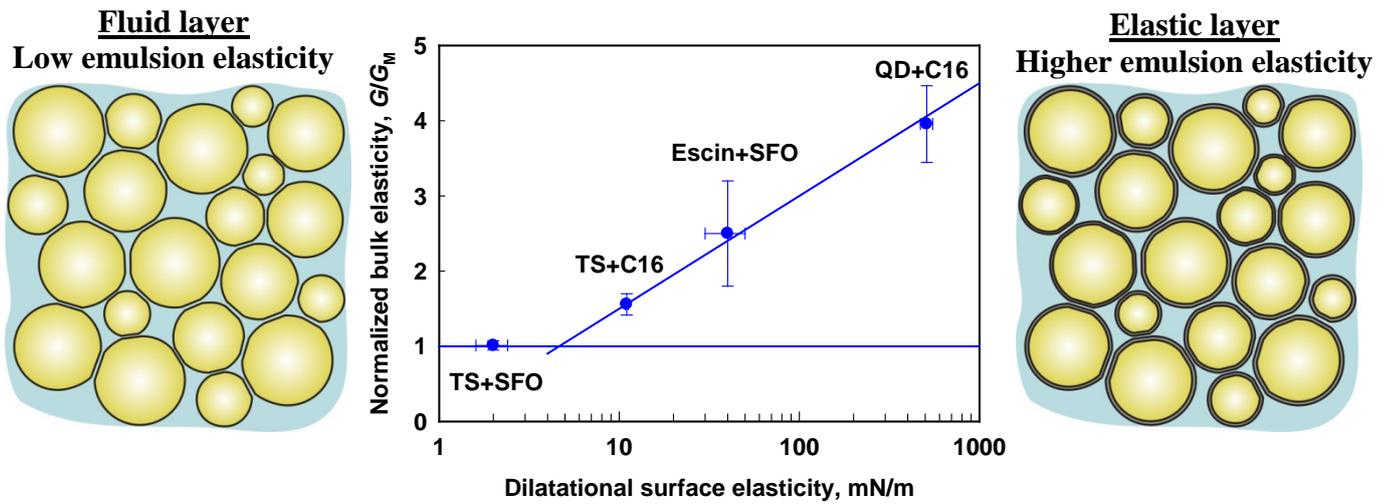

# Supporting information

**Table S1.** Oils used in the study.

| Type of oil | Abbreviation used in the text | Purity, % | $\sigma_0$, mN/m | Density, g/ml | Viscosity, mPa.s | Producer |
|---|---|---|---|---|---|---|
| Hexadecane | C16 | 99 % | 52.5 | 0.77 | 3.45 (20 °C) | Sigma Aldrich |
| Sunflower oil | SFO | - | 30.5 | 0.92 | 49.14 (25 °C) [27] | - |

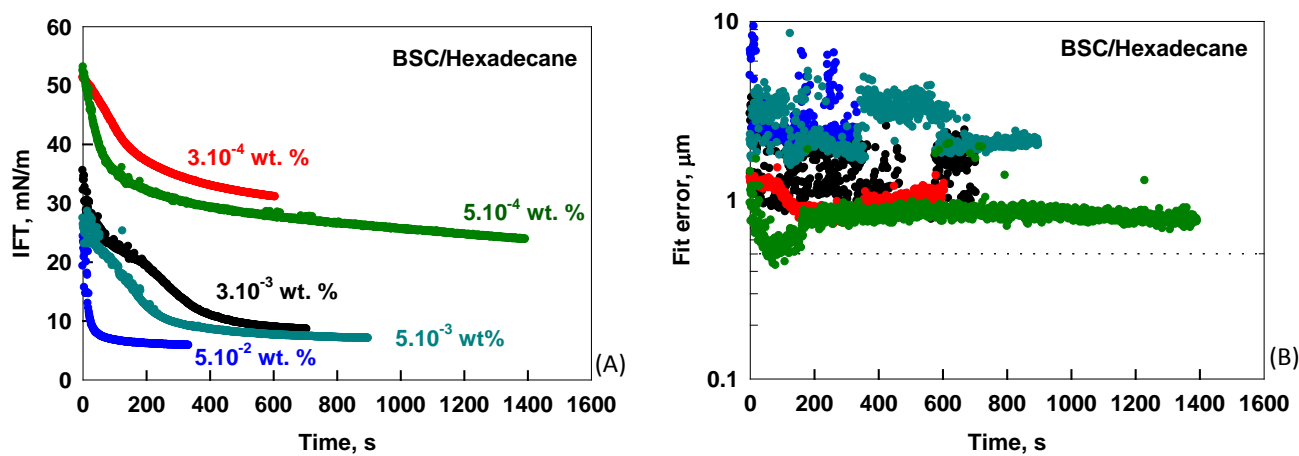

**Figure S1.** Analysis of the shape of the drops (DSA) of saponin solutions in hexadecane for BSC solutions with different concentrations, as indicated in the figure. (A) Interfacial tension vs. time; (B) Fit error of the drop shape vs. time. The dashed line indicates the typical value of the Fit error for fluid layers.

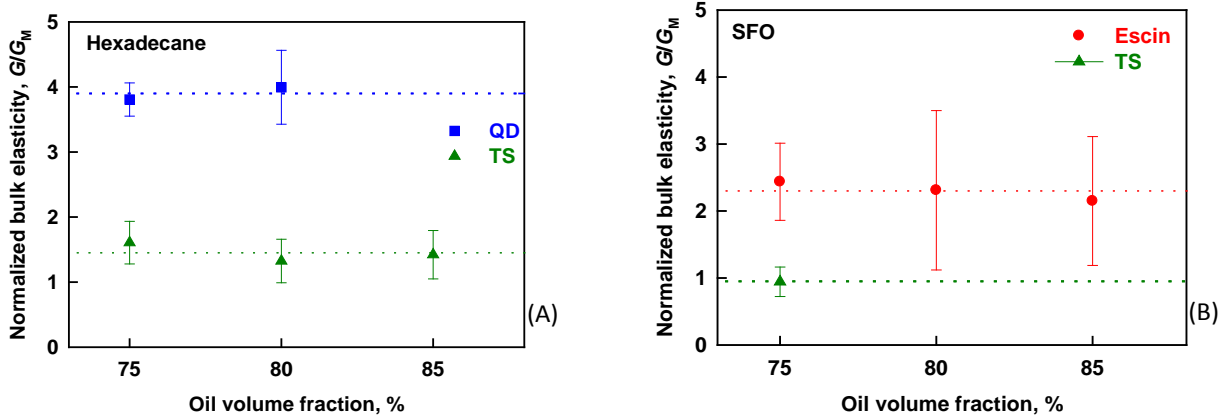

**Figure S2.** Emulsion elastic modulus as measured in our experiments divided by the emulsion elastic modulus predicted by eq. (7) as a function of oil volume fraction for (A) hexadecane and (B) SFO emulsions stabilized by TS (green symbols); Escin (red circles) and QD (blue squares).

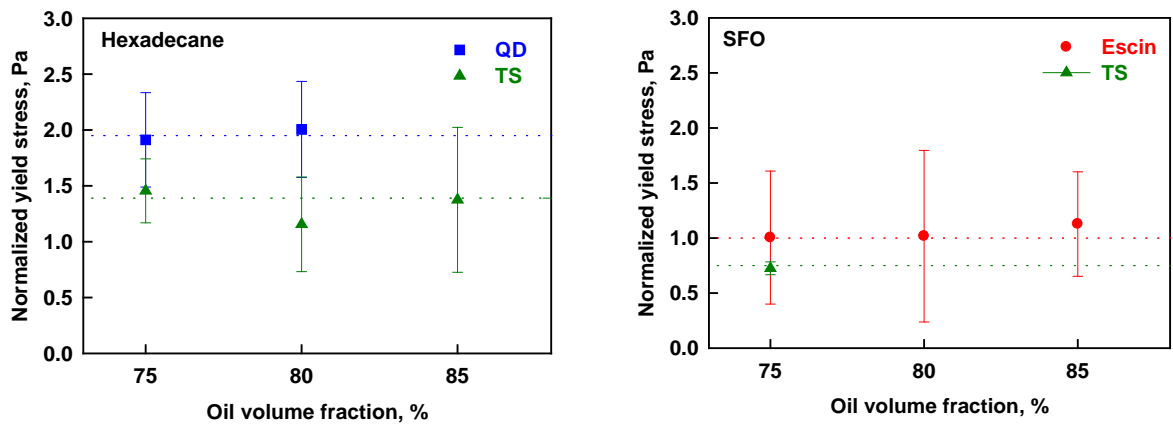

**Figure S3.** Yield stress as measured in our experiments, divided by the yield stress predicted by eq. (9) as a function of the oil volume fraction for: (A) hexadecane and (B) SFO emulsions stabilized by TS (green symbols); Escin (red circles) and QD (blue squares).